# Effects of the Hunga Tonga–Hunga Ha'apai Volcanic Eruption on Observations at Paranal Observatory

Robert J. De Rosa [1]
Angel Otarola [1]
Thomas Szeifert [1]
Jonathan Smoker [1,2]
Fernando Selman [1]
Andrea Mehner [1]
Fuyan Bian [1]
Elyar Sedaghati [1]
Julia Victoria Seidel [1]
Alain Smette [1]
Willem-Jan de Wit [1]

[1] ESO
[2] UK Astronomy Technology Centre, Royal Observatory, Edinburgh, UK

**The Hunga Tonga–Hunga Ha'apai volcano erupted on 15 January 2022 with an energy equivalent to around 61 megatons of TNT. The explosion was bigger than any other volcanic eruption so far in the 21st century. Huge quantities of particles, including dust and water vapour, were released into the atmosphere. We present the results of a preliminary study of the effects of the explosion on observations taken at Paranal Observatory using a range of instruments. These effects were not immediately transitory in nature, and a year later stunning sunsets are still being seen at Paranal.**

## Introduction

Many astronomers were likely unaware of the impact of volcanic eruptions on astronomical observations when the Hunga Tonga–Hunga Ha'apai (HTHH) volcano erupted in January 2022. The release of sulphur dioxide ($SO_2$) by such eruptions can have a significant effect on the atmospheric transmission and on Earth's climate. When injected into the stratosphere, $SO_2$ is transformed into sulphuric acid ($H_2SO_4$) by the photochemical effect in the presence of water vapour. While Earth's surface effectively cools because the Sun's short wavelength radiation is scattered by the sulphate aerosols in the stratosphere, heat radiation from Earth's surface is efficiently absorbed by the same particles, resulting in an altered weather pattern and climate. The sulphate aerosol particles in the stratosphere circulate globally and are only removed by precipitation on timescales of several years.

In terms of the energy released, the world's biggest volcanic incident in the past 1300 years was the eruption of Mount Tambora, in what is now Indonesia, in April 1815. In addition to the immediate and direct destruction, a huge dust cloud entered the stratosphere, which disrupted weather systems in 1816 and for the following three years in the northern hemisphere. The year 1816 was the second-coldest year on record since the Middle Ages and is known as the 'year without a summer'. The change in climate was followed by famine, disease, poverty and civil unrest, with many social and political consequences (D'Arcy Wood, 2014; Behringer, 2019). The years after the Tambora eruption also sparked the imaginations of artists (for example Mary Shelley's Frankenstein, the dark poetry of Lord Byron, paintings by J. M. W. Turner and Caspar David Friedrich, and music by Beethoven and Schubert)[1,2]. Interestingly, the link between climate change in the 1810s and the volcanic eruption of Tambora was not recognised at the time. This connection was only realised after the eruption of the Krakatoa in Indonesia in 1883 (Royal Society, 1888), at a time when news could be quickly reported via telegraphy across the world.

## The eruption of the Hunga Tonga–Hunga Ha'apai volcano

The submarine HTHH volcano in the South Pacific erupted violently on 15 January 2022. An ash plume shot 57 km into the mesosphere, shockwaves rippled through the atmosphere, and the eruption triggered a tsunami with heights of more than 19 m above sea level[3], causing massive infrastructure destruction on the nearby islands and the death of four people in Tonga and two in Peru. The energetic output from the volcano has been estimated to be approximately 61 megatons of TNT equivalent (Diaz & Rigby, 2022). Ocean floor maps showed that the volcano spewed out at least 9.5 $km^3$ of material in total[4]. By comparison, the 1815 eruption of Tambora in Indonesia ejected more than 100 $km^3$ of erupted material, the 1883 eruption of Krakatoa in Indonesia 25 $km^3$, the 1991 eruption at Mount Pinatubo in the Philippines 5.5 $km^3$, and the CE 79 eruption of Mount Vesuvius 4 $km^3$. In the case of HTHH, 1.9 $km^3$ of material ended up in the atmosphere, which caused the stunning sunsets observed from Paranal following the eruption.

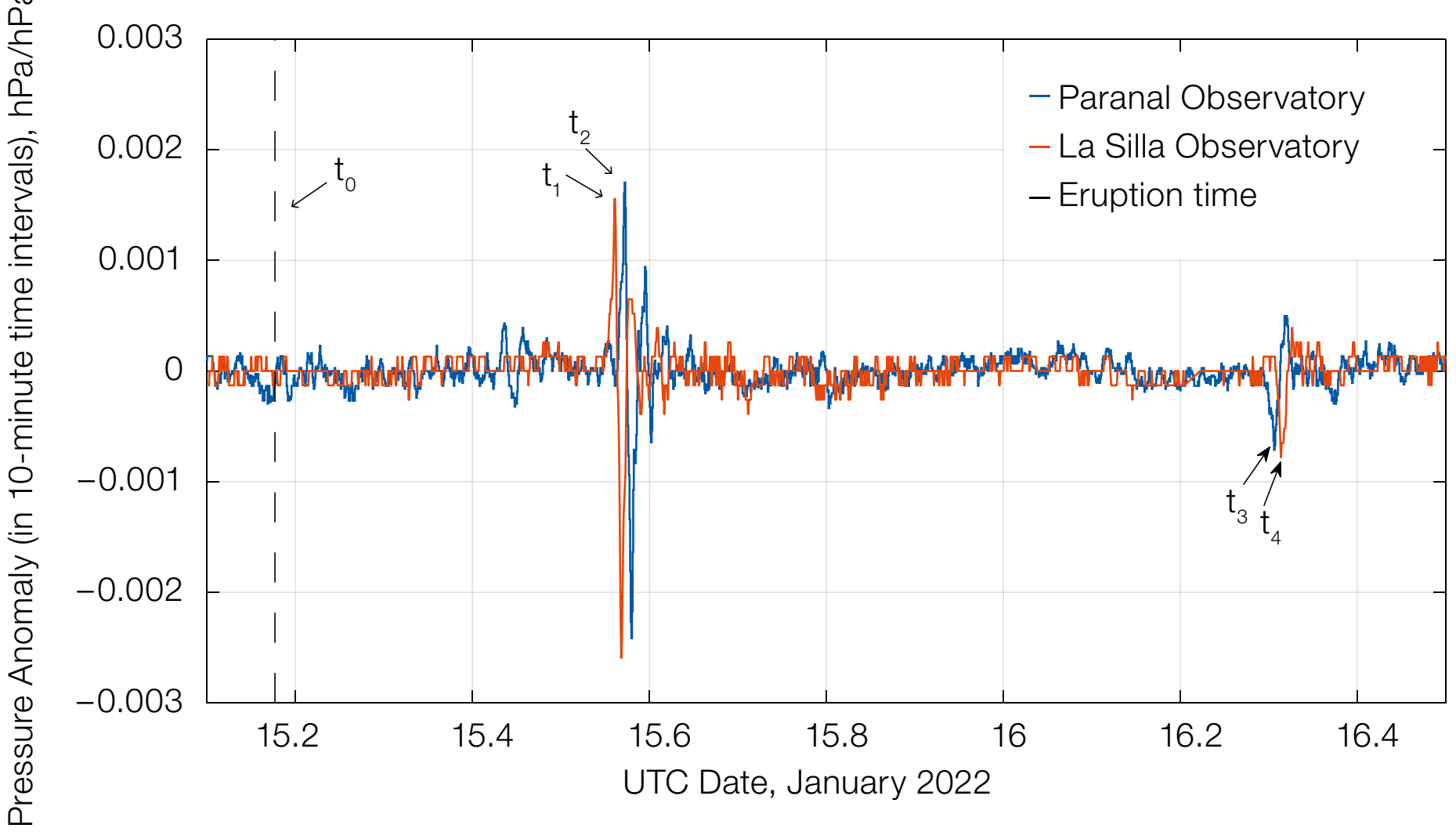

Figure 1. Atmospheric pressure anomaly. Observations on Paranal and La Silla of the pressure wave caused by the HTHH volcanic eruption. The indicated time stamps are: $t_0$ = eruption moment (2022-01-15 04:14:45 UT), $t_1$ = arrival of the front edge of the pressure wave at La Silla Observatory (2022-01-15 13:28:42 UT), $t_2$ = arrival of the front edge of the pressure wave at Paranal Observatory (2022-01-15 13:43:49 UT), $t_3$ = arrival of the far edge of the pressure wave at Paranal Observatory (2022-01-16 07:45:32 UT), $t_4$ = arrival of the far edge of the pressure wave at La Silla Observatory (2022-01-16 07:20:13 UT).

Conclusions drawn in the work of Legras et al. (2022) and references therein describe the following picture that characterises the event: a) the eruption was intense with a large injection of water vapour into the stratosphere; b) the zonal atmospheric circulation spread the plume across all longitudes in less than one month; c) after six months the plume, mainly sulphates and water vapour, had spread in the 35°S–20°N latitude range, in two plumes separated in latitude; d) satellite observations of the plume in the optical and millimetre spectral bands and the observed sedimentation rate seem to show the sulphate aerosol particles reached a size of about 1.0 and 1.4 µm in diameter. The size evolution of the particles has been explained by hygroscopic growth during the first phase (up to about April), followed by coagulation (in the period April–May) and then decay by evaporation in the later stage which is dominated by evaporation as a result of dry air and a more diluted plume.

### Detection of the atmospheric shockwave at La Silla and Paranal Observatories

The shockwave from the eruption was detected at various weather stations around the world (Harrison, 2022). Figure 1 shows the pressure anomaly caused by the shockwave passing over both Paranal and La Silla in Chile, a distance exceeding 10 000 km from the eruption. From the geodetic distances between the volcano site and the ESO observatories, and the elapsed times for the arrival of the atmospheric pressure wavefronts we were able to compute the average speed of the pressure wave to be approximately 307 m $s^{-1}$. Moreover, the ESO weather stations show both the arrival of the front edge and the far edge of the pressure wave. The front edge is the part of the shockwave that propagated from the location of the volcano towards the East in the direction of Chile. The far edge is the part of the shockwave that propagated in the opposite direction, around the planet, before reaching the observatory sites. The amplitude of the pressure anomaly for the far-edge wave is down to 30% of that of the front-edge wave, an indication of the loss of energy as the shockwave propagated away from the main event.

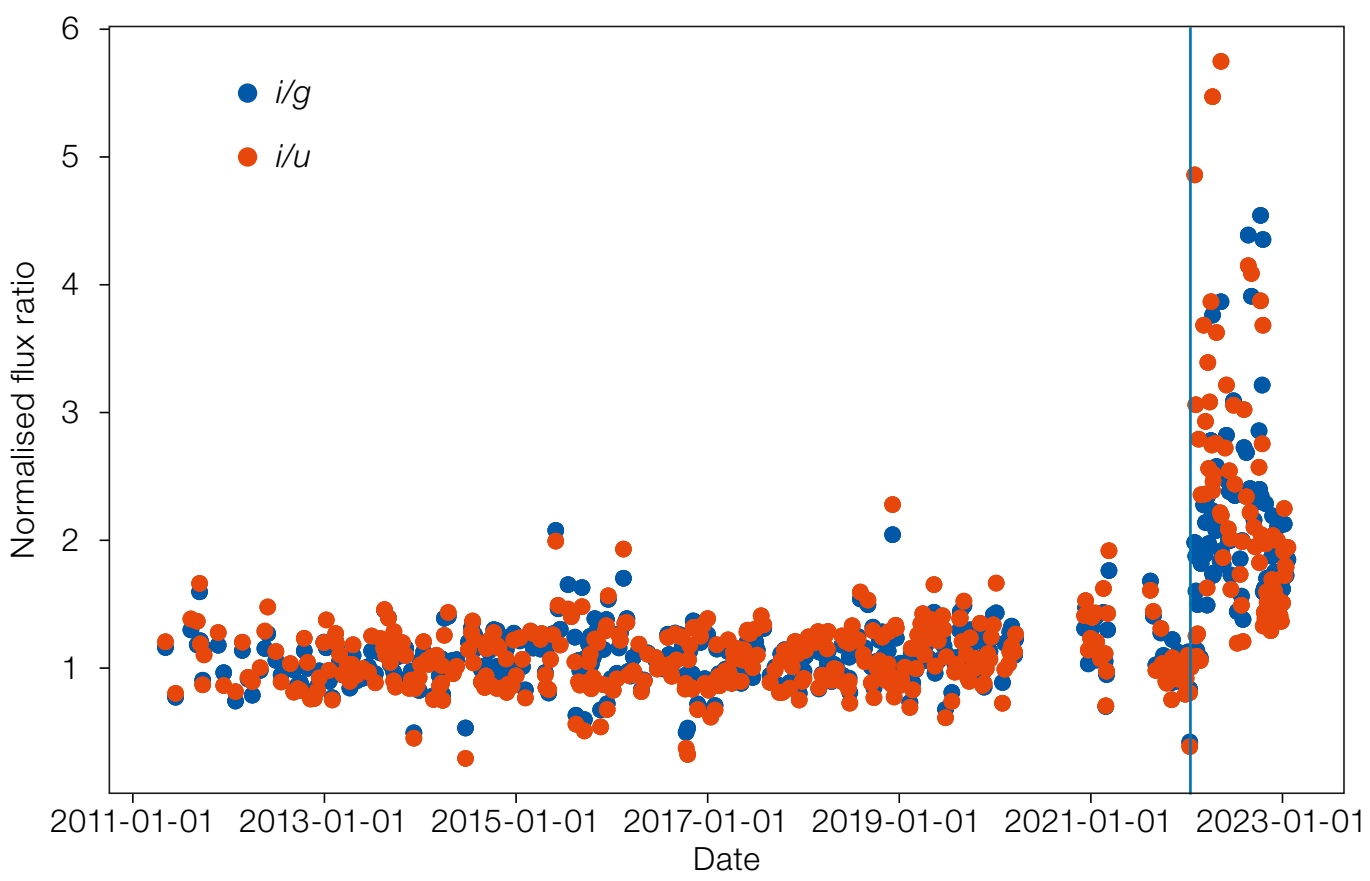

**Figure 2.** Flux ratio for *i/g* (blue symbols) and *i/u* (red symbols) sky flatfields taken with OmegaCAM at the VLT Survey Telescope as a function of time. Observations in the different bands were taken simultaneously with the mosaic filter and are normalised to the pre-explosion ratio. The vertical line indicates the date of the volcanic eruption.

## Effects on Paranal

### Sky flatfields

Particles of volcanic dust in the atmosphere can cause spectacular sunsets, which have been captured in paintings, for example by the artists J. M. W. Turner and Caspar David Friedrich, after the Tambora eruption in 1815. Following the eruption of the HTHH volcano, there was some awareness of this effect on Paranal, because of publications of several post-eruption standard-star observations on La Silla (Rufener, 1986a,b; Grothues & Gochermann, 1992; Burki et al., 1995a,b). In the aftermath of the HTHH eruption, the colour of the Paranal sunset was very different from what we were used to seeing in the previous years.

During normal operations at Paranal, sky flatfields are taken on a regular basis to calibrate scientific data and to monitor the instrument health. Figure 2 shows the normalised flux ratio against time for optical flatfields taken at twilight using the segmented filter of the OmegaCAM camera at the Very Large Telescope (VLT) Survey Telescope, where images are taken simultaneously with the *ugri* SDSS filters. A few days after the volcanic eruption, the ratio of the *i* (770 nm) and both the *u* (350 nm) and *g* (480 nm) flatfields increased by a factor of five, indicating a significant reddening of the twilight sky, and these have still not returned to the pre-explosion values one year later. These measurements are consistent with perceived changes of the colour of sunsets seen at Paranal. Dome flatfields obtained with the same filters do not show any change in the flux ratio, indicating that the observed effect is atmospheric and not instrumental.

A similar effect was seen in the sky flats taken with the VIRCAM camera at the Visible and Infrared Survey Telescope for Astronomy (VISTA) where the ratio of the $K_s$ (2146 nm) to $Y$ (1021 nm) twilight flats increased after the explosion. A sudden change in the decay time of the twilight was also observed in near-infrared data from the HAWK-I instrument at the VLT. Figure 3 shows the count rate during twilight with the $K_s$ filter as a function of the elevation of the Sun, before and after the volcanic eruption.

### Model of the observed changes in the HAWK-I twilight sky brightness

The change in the decay time of the sky brightness in the HAWK-I twilight flatfields can be explained by a 36-km-high column of dust in the line of site of the telescope, which became the dominant source of reflection rather than the molecular gas in the troposphere before the eruption. The decay of the reflected light observed in the twilight sky flats can be explained by the rising of the Earth shadow, as illustrated in Figure 4, where the Rayleigh scattering region is shown in blue, and the stratospheric layer with the red circle. Twilight ends when the Earth shadow reaches the upper boundary of the stratospheric layer of dust. From this, the height of the volcanic dust plume can be estimated as 36 km in March 2022 compared to the 57-km height of the dust dome above the volcano immediately after outburst (Proud, Prata & Schmauß, 2022). In observations obtained in November 2022, the flatter twilight flat-field decay time indicates a decreasing density of the dust in the stratosphere and a reduction in the height of the dust layer down to 29 km, indicated by the later onset of the infrared twilight and illustrated in Figure 3.

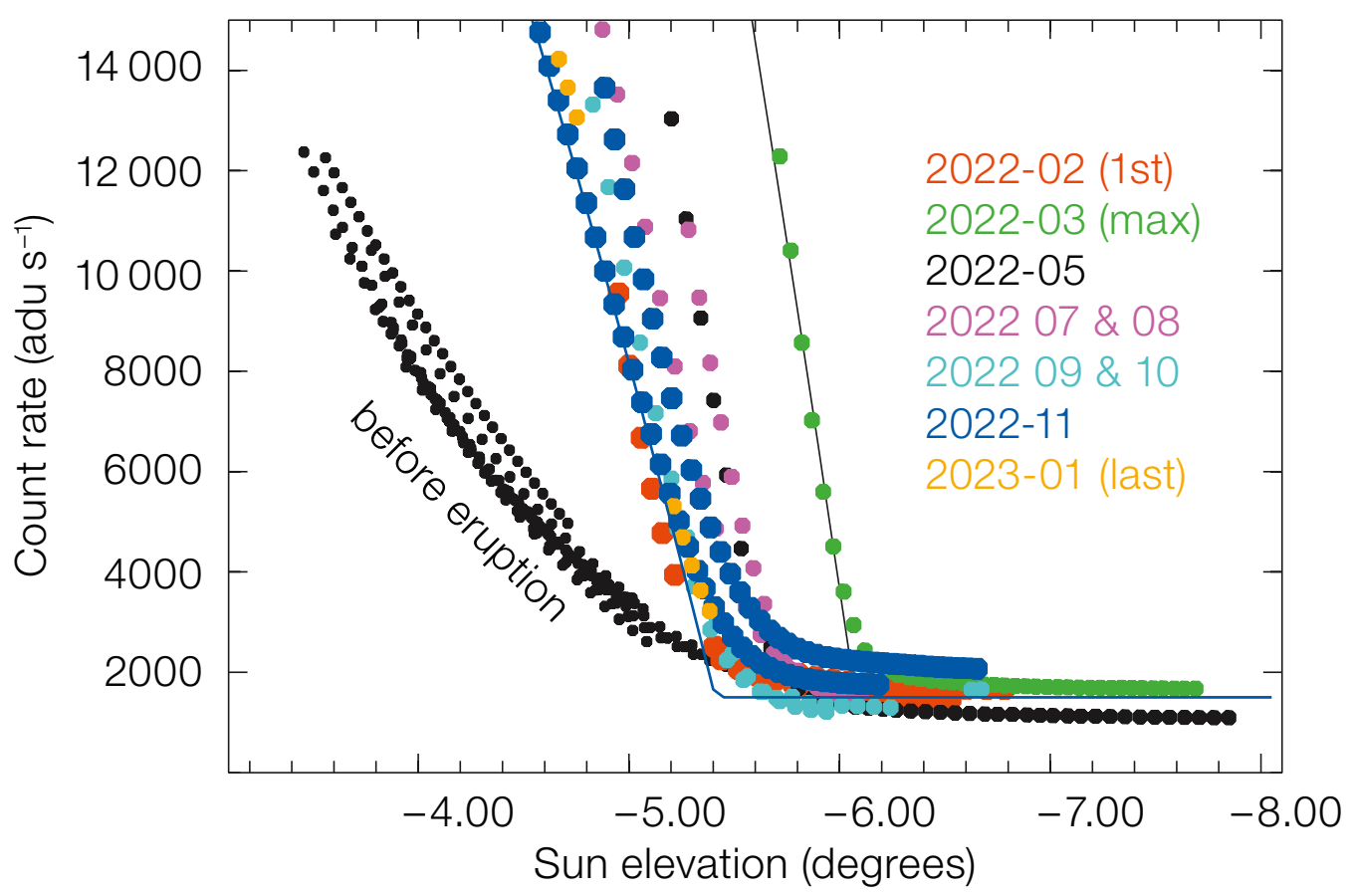

**Figure 3.** Decay of the twilight flux in near-infrared light ($K_s$) observed with HAWK-I as a function of the elevation of the Sun below the horizon. The green line is a model for March 2022, which estimates a column height of the volcanic dust plume of 36 km. The blue line indicates a model for November 2022, which estimates a height of 29 km and 0.6 times less reflected light per meter of illuminated column height.

## Atmospheric extinction and zero points

The atmospheric extinction is an important parameter for photometric measurements. Besides seasonal and other long-term variations, significant increases of the extinction coefficients can be caused by major volcanic eruptions that inject large amounts of aerosols into the stratosphere at altitudes between 20 and 30 km. These aerosols are distributed over wide areas of Earth's atmosphere by jetstreams and can influence astronomical observations at great distances (for example, Moreno & Stock, 1964; Rufener, 1986a,b; Grothues & Gochermann, 1992; Burki et al., 1995a,b).

Long-term variations of the extinction caused by the volcanic eruptions of El Chichón in Mexico in 1982 and the Pinatubo eruption in the Philippines in 1991 were observed extensively at La Silla (Rufener, 1986a,b; Grothues & Gochermann, 1992; Burki et al., 1995a,b). The stratospheric load due to the eruption of El Chichón was estimated to be about 8 megatons of $SO_2$ and the Pinatubo eruption emitted about 20 megatons of $SO_2$ into the atmosphere. The increase of the extinction in both events was very sudden — roughly 150 days (El Chichón) and 100 days (Pinatubo) after the eruptions. The removal of the volcanic aerosols from the atmosphere lasted at least 1000 days. Aerosols from different volcanoes are very different; for example, Pinatubo produced a flatter extinction curve than El Chichón.

To search for the possible effect of the aerosols injected into the upper atmosphere by the eruption of the HTHH volcano, we analysed the extinction measurements obtained from the observation of standard stars utilising the FORS2 imager at the VLT on Paranal. Data from the wide-field imager OmegaCAM were also included in the analysis, although in this case we do not measure the atmospheric extinction, assuming instead a constant value. Hence, a variation in extinction is observed as a change in zero point which we monitor for each detector and filter; an increase in atmospheric extinction should appear as a reduction in the zero point of similar magnitude.

The corrected extinction data for four FORS2 filters and only for stable night transparency conditions are plotted in Figure 5. They were obtained as a part of the QC1 quality control process[5] following the method developed for the FORS Absolute Photometry project[6], as described by Freudling et al. (2007). The data, for filters $b_{HIGH}$ (440 nm), $v_{HIGH}$ (557 nm), $R_{SPECIAL}$ (655 nm) and $I_{Bessel}$ (768 nm), cover the period from 1 January 2021 until the end of December 2022. The date of the eruption is shown with a vertical line. The increase in extinction that could be attributed to an increase in aerosol content in the stratosphere (at 25–27 km altitude) from the volcano is not significant compared to the seasonal variability in the sky extinction. The seasonal variability is due to the change in the atmospheric absolute humidity and barely visible cirrus condition induced by the altiplanic winter. The OmegaCAM data show an analogous seasonal variability.

## Water vapour

The HTHH explosion released substantial amounts of water vapour into the stratosphere (Millán et al., 2022; Legras et al., 2022). The work of Legras et al. (2022), with data from the satellite-borne Microwave Limb Sounder (MLS) instrument, shows the eruption of the volcano injected a plume of water vapour of up to 25 ppmv concentration in the layer between 20 and 30 km above sea level. This is five times more than the background water vapour level at that altitude for the tropical latitude of the Paranal Observatory.

At the time of the volcanic eruption, two water vapour radiometers were in operation at Paranal (LHATPRO; Kerber et al., 2012). Unfortunately, the increase in water vapour at such high altitudes is not detectable at ground level using the LHATPRO. This is because the water vapour emission in the stratosphere gets attenuated by absorption along the long path to the LHATPRO detector on Earth's surface. We computed the atmospheric brightness temperature using the atmospheric model *am* (Paine, 2022) and find that a change in brightness temperature at the peak of the 183 GHz water vapour line induced by a 25 ppmv water vapour layer, with respect to the background condition, is only around 0.07%, corresponding to a change of 0.04% in precipitable water vapour, which is well below the daily temporal variability of water vapour at Paranal.

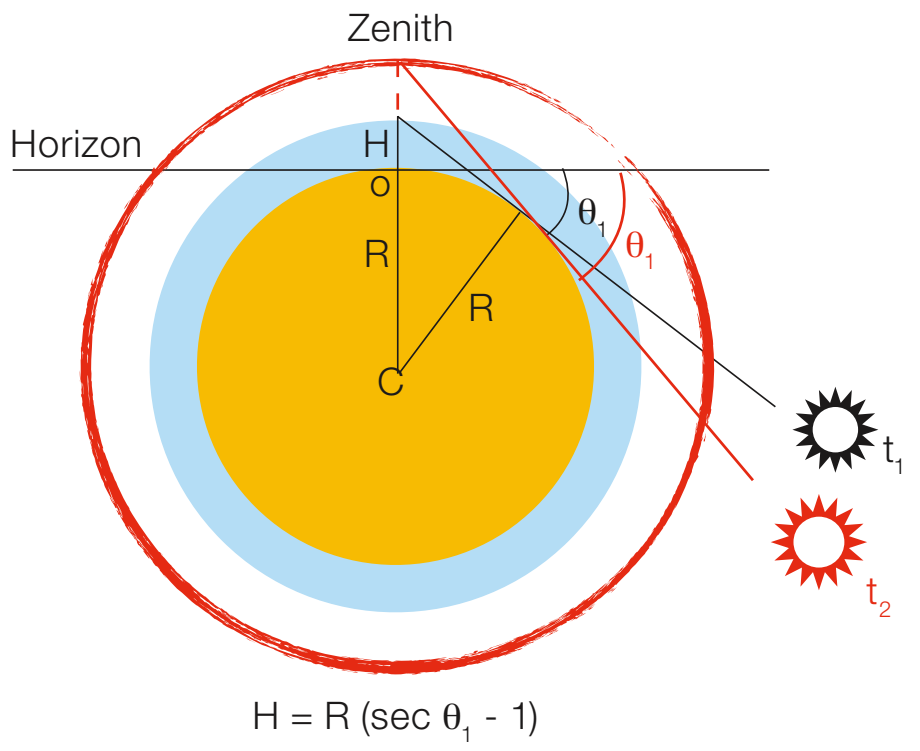

**Figure 4.** Schematic diagram (not to scale) of our model to explain the sky brightness variations during twilight caused by scattering from aerosols at different altitudes within the stratosphere. The Sun at time $t_2$ only illuminates the higher regions in the stratosphere when observing the zenith from position O.

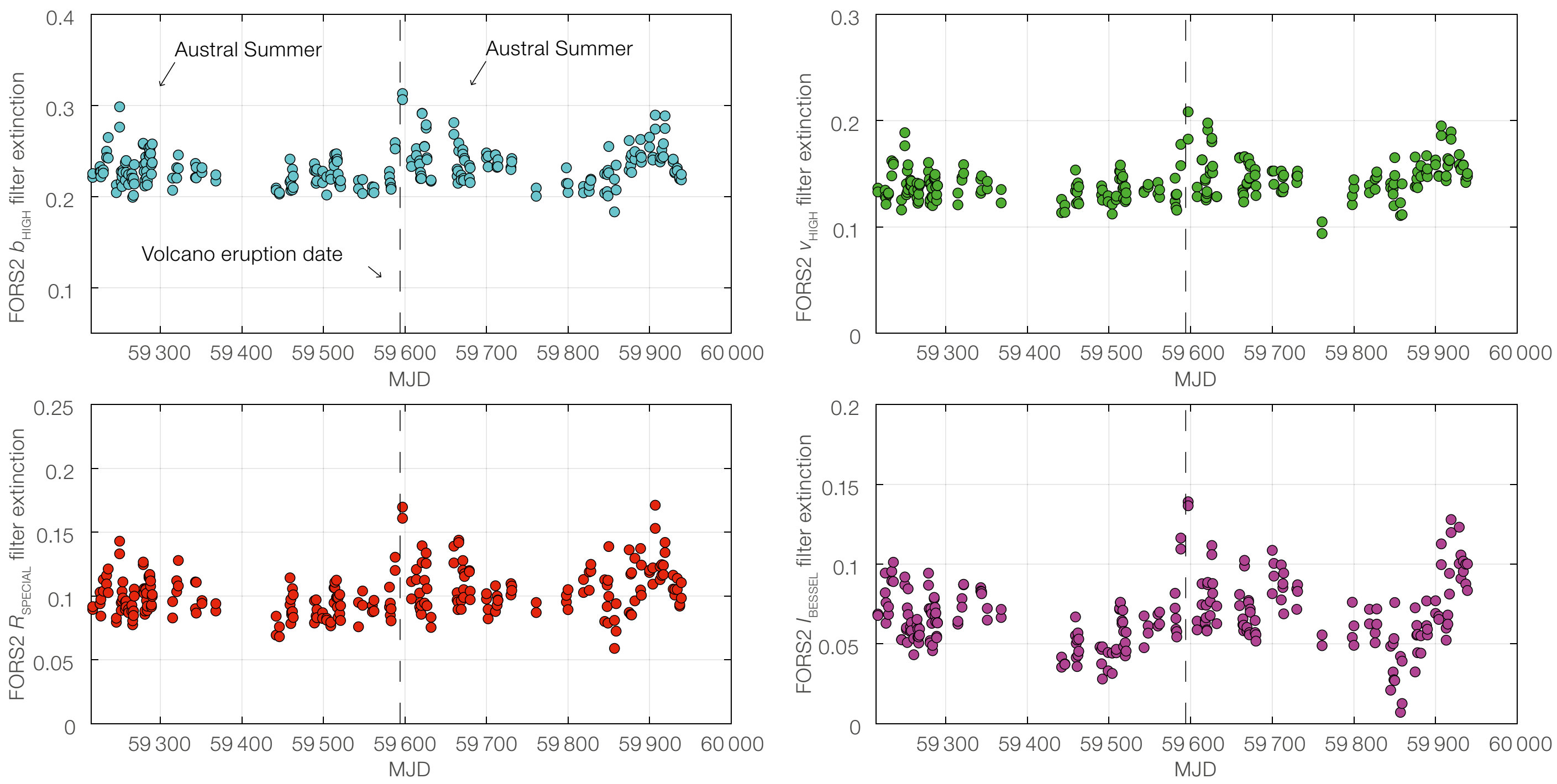

**Figure 5.** FORS2 extinction time series[5] between 1 January 2021 and 30 December 2022. The vertical line in each subplot shows the date of the eruption. (top-left) $b_{HIGH}$ filter, (top-right) $v_{HIGH}$ filter, (bottom-left) $R_{SPECIAL}$ filter, (bottom-right) $I_{BESSEL}$ filter.

## Conclusions

The HTHH volcanic eruption had several observable effects at Paranal Observatory, located more than 10 000 km away from the eruption site. These included measurements of the shockwave and a striking change in the colour of the sky visible in routine calibration sky flats taken during twilight at optical and infrared wavelengths. The detection of variations in the extinction, observed via standard star observations, is harder to assess since the volcanic eruption took place during the austral summer. This is a period when we observe a natural variability and increase of the atmospheric extinction owing to an increase in humidity and barely visible cirrus, induced by the altiplanic winter atmospheric condition. The detection of the increase in water vapour, injected by the volcano in the stratospheric level, was not possible because of the low sensitivity of our water vapour radiometers to emission from such a high altitude. A more detailed analysis of the data is ongoing to determine the composition of the aerosols and to monitor the longer-term effects of the explosion on astronomical observations. Given the intensity of the eruption and the large amount of water vapour injected in the stratosphere, it is believed the effects will last for several years.

## References

Behringer, W. 2019, *Tambora and the Year without a Summer: How a Volcano Plunged the World into Crisis* (Medford, MA: Polity Press)
Burki, G. et al. 1995a, The Messenger, 80, 34
Burki, G. et al. 1995b, A&AS, 112, 383
D'Arcy Wood, G. 2014, *Tambora: The Eruption That Changed the World* (Princeton: Princeton University Press)
Diaz, J. S. & Rigby, S. E. 2022, Shock Waves, 32, 553
Freudling, W. et al. 2007, The Messenger, 128, 13
Grothues, H.-G. & Gochermann, J. 1992, The Messenger, 68, 43
Harrison, G. 2022, Weather, 77, 87
Kerber, F. et al. 2012, Proc. SPIE, 8446, 84463N
Legras, B. et al. 2022, Atmos. Chem. Phys., 22, 14957
Millán, L. et al. 2022, Geophys. Res. Lett., 49, 99381
Moreno, H. & Stock, J. 1964, PASP, 76, 55
Proud, S. R., Prata, A. T. & Schmauß, S. 2022, Science, 378, 554
Royal Society 1888, *The Eruption of Krakatoa: And Subsequent Phenomena* (London, Trübner & Company)
Rufener, F. 1986a, The Messenger, 44, 32
Rufener, F. 1986b, A&A, 165, 275
Paine, S. 2022, The am atmopsheric model (v. 12.2), https://doi.org/10.5281/zenodo.6774376

## Links

[1] Ian Ritchie's website: https://www.ianritchie.org/the-year-without-a-summer
[2] The Guardian website: https://www.theguardian.com/music/2016/jun/16/1816-year-without-summer-dark-masterpieces-beethoven-schubert-shelley
[3] The BBC website: https://www.bbc.com/news/science-environment-61567521
[4] National Institute of Water and Atmospheric Research website: https://niwa.co.nz/news/tonga-eruption-confirmed-as-largest-ever-recorded
[5] FORS2 QC1 processed zeropoint and extinction data: http://archive.eso.org/qc1/qc1_cgi?action=qc1_browse_table&table=fors2_photometry
[6] See the section FORS2 *Absolute Photometry Project* at https://www.eso.org/sci/facilities/paranal/instruments/fors/doc.html